\documentclass[aps,10pt,twocolumn,superscriptaddress,footinbib,flushbottom]{revtex4-1}%

\usepackage{amssymb,amsmath,amsfonts,bm,dcolumn}
\usepackage{graphicx,color}
\usepackage[squaren]{SIunits}
\usepackage[english]{babel}
\usepackage{tabularx}
\usepackage{tikz}

\renewcommand{\vec}[1]{\boldsymbol{\mathrm{#1}}}%
\newcommand{\uvec}[1]{\bm{\hat{\mathbf{#1}}}}

\newcommand{\mean}[1]{\langle #1 \rangle}

\renewcommand{\O}[1]{\mathcal{O}\left(#1\right)}

\newcommand{\Nabla}{\vec{\nabla}}%

\newcommand{\dif}{d}

\newcommand{\integral}[4]{\int_{#1}^{#2} \! #3 \, \dif#4}%

\newcommand{\fct}[1]{\mathcal{#1}}

\newcommand*\showwidth[1]{%
	\textcolor{blue}{\rule{\csname#1\endcsname}{1pt}}\newline
	\texttt{\textbackslash#1}: \expandafter\the\csname#1\endcsname
	\par
}

\newmuskip\pFqmuskip
\newcommand*\pFq[6][8]{%
	\begingroup 
	\pFqmuskip=#1mu\relax
	\mathchardef\normalcomma=\mathcode`,
	\mathcode`\,=\string"8000
	\begingroup\lccode`\~=`\,
	\lowercase{\endgroup\let~}\pFqcomma
	{}_{#2}F_{#3}{\left[\genfrac..{0pt}{}{#4}{#5};#6\right]}%
	\endgroup
}
\newcommand{\pFqcomma}{{\normalcomma}\mskip\pFqmuskip}

\graphicspath{ {figure/} }

\begin{document}
	
	\title{Dynamics of active particles with translational and rotational inertia}

	\author{Alexander R. Sprenger}
    \email{alexander.sprenger@hhu.de}
	\affiliation{Institut f\"{u}r Theoretische Physik II: Weiche Materie, Heinrich-Heine-Universit\"{a}t D\"{u}sseldorf, D-40225 D\"{u}sseldorf, Germany}
	\affiliation{Institut für Physik, Otto-von-Guericke-Universit\"at Magdeburg, Universit\"atsplatz 2, D-39106 Magdeburg, Germany}
 
	\author{Lorenzo Caprini}
    \email{lorenzo.caprini@gssi.it}
	\affiliation{Institut f\"{u}r Theoretische Physik II: Weiche Materie, Heinrich-Heine-Universit\"{a}t D\"{u}sseldorf, D-40225 D\"{u}sseldorf, Germany}

	\author{Hartmut L\"{o}wen}
	\affiliation{Institut f\"{u}r Theoretische Physik II: Weiche Materie, Heinrich-Heine-Universit\"{a}t D\"{u}sseldorf, D-40225 D\"{u}sseldorf, Germany}
	
	\author{Ren\'e Wittmann}
	\affiliation{Institut f\"{u}r Theoretische Physik II: Weiche Materie, Heinrich-Heine-Universit\"{a}t D\"{u}sseldorf, D-40225 D\"{u}sseldorf, Germany}

	\date{\today}
	
	\begin{abstract}
	    Inertial effects affecting both the translational and rotational dynamics are inherent to a broad range of active systems at the macroscopic scale.
	    Thus, there is a pivotal need for proper models in the framework of active matter to correctly reproduce experimental results, hopefully achieving theoretical insights.
	    For this purpose, we propose an inertial version of the active Ornstein-Uhlenbeck particle (AOUP) model accounting for particle mass (translational inertia) as well as its moment of inertia (rotational inertia) and derive the full expression for its steady-state properties. 
		The inertial AOUP dynamics introduced in this paper is designed to capture the basic features
		of the well-established inertial active Brownian particle (ABP) model, i.e., the persistence time of the active motion and the long-time diffusion coefficient. 
		For a small or moderate rotational inertia, these two models predict similar dynamics at all timescales and, in general, our inertial AOUP model consistently yields the same trend upon changing the moment of inertia for various dynamical correlation functions.
	\end{abstract}
	
	\maketitle

	\section{Introduction}
	Active motion can be observed at both microscopic and macroscopic scales~\cite{marchetti2013hydrodynamics, elgeti2015physics, bechinger2016active}, with typical examples ranging from birds, fish and insects to colloids and bacteria or cell monolayers. 
	A common feature of such active systems is the capability to convert energy from the environment to produce directed motion~\cite{bechinger2016active, gompper20202020}, which allows them to swim, move or fly in their environment.
	As a consequence, their dynamics qualitatively differs from that of ''passive'' Brownian particles, originally introduced to describe the random motion of pollen grains in water solution~\cite{gardiner1985handbook} and extensively employed to model colloidal particles.
	While the (overdamped) passive motion of passive colloids is characterized by random (Brownian) trajectories, showing a pure diffusive behavior,  
	active motion generally gives rise to persistent single-particle trajectories~\cite{bechinger2016active, shaebani2020computational}: an active particle typically moves persistently in one spatial direction with a typical velocity, known as the swim velocity, and only after a typical time, known as persistence time, randomizes its direction of motion.
	
	These features have been identified as the basic ingredients to build coarse-grained models in the framework of stochastic processes, able to capture the essential behavior of this class of active systems. 
	Among them, the famous model of active Brownian particles (ABPs)~\cite{ten2011brownian, sevilla2015smoluchowski, solon2015pressure, petrelli2018active, caprini2020hidden, shi2020self, gomez2020active, martin2021characterization,sprenger2022active} introduces the ''activity'' as a time-dependent force of constant magnitude with a stochastic evolution of its direction. 
    It is commonly used due to its simplicity while it also presents an accurate representation of active colloids~\cite{buttinoni2013dynamical, palacci2013living, aranson2013active, takatori2016acoustic, zottl2016emergent} subject to both translational and rotational Brownian motion.
	Recently, an alternative model, known as active Ornstein-Uhlenbeck particles (AOUPs)~\cite{martin2021statistical, berthier2019glassy, dabelow2019irreversibility, sevilla2019generalized, woillez2020nonlocal}, has been introduced, firstly, to describe the motion of a passive colloid in a bath formed by active bacteria~\cite{wu2000particle, maggi2014generalized, maggi2017memory, maes2020fluctuating}, and, secondly, to further simplify the ABP dynamics in terms of Gaussian correlations~\cite{fily2012athermal, szamel2014self, farage2015effective}, which allows to obtain exact analytical predictions~\cite{maggi2015multidimensional, fodor2016far, martin2021aoup, caprini2021fluctuation} or devise approximate theories~\cite{marconi2016velocity,sharma2017escape,wittmann2017effective}.
	The two models show consistent results, being both able to reproduce the typical non-equilibrium phase coexistence of active particles, known as motility-induced phase separation (MIPS)~\cite{fily2012athermal, buttinoni2013dynamical, speck2014effective, digregorio2018full, cates2015motility, caporusso2020motility, martin2021characterization, caprini2020spontaneous, PhysRevLett.129.048002}, as well as the accumulation or wetting at boundaries or generic obstacles~\cite{li2009accumulation,ni2015tunable,wittmann2016active,wittmann2019pressure,turci2021wetting}.
	Beyond the qualitative level, the results of the two models have been compared in several cases of interest~\cite{farage2015effective, das2018confined, caprini2019comparative}, and, recently, their relation has been comprehensively investigated in Ref.~\cite{caprini2022parental}.

	Both ABPs and AOUPs have been originally developed to model the overdamped dynamics of microscopic active particles.
	However,  also macroscopic active ``particles'' are rather common in the animal world, such as birds~\cite{cavagna2014bird}, fish \cite{pavlov2000patterns} and insects~\cite{mukundarajan2016surface,FeinermanPGFG2018}, as well as in the inanimate world, such as walking droplets~\cite{valani2019superwalking}, flying whirling fruits~\cite{rabault2019curving} and active granular particles~\cite{weber2013long, koumakis2016mechanism, scholz2018rotating, dauchot2019dynamics, leoni2020surfing, soni2020phases, baconnier2021selective}.
	The recent significant increase of interest in these systems generates the need to develop manageable generalized theoretical descriptions including inertial effects~\cite{lowen2020inertial}.

	The first, and most obvious, step to model inertial active systems, is to account for a larger particle's mass or, equivalently, a smaller translational friction coefficient.
	Such inertial forces are easily included in an underdamped description for the translational motion of ABPs~\cite{joyeux2016pressure, ai2017transport, shankar2018hidden, mandal2019motility, wagner2019response, vuijk2020lorentz, holubec2020underdamped, martins2022inertial} and AOUPs~\cite{cecconi2018anomalous, lee2019thermodynamic, caprini2020inertial, nguyen2021active, goswami2022inertial, frydel2022entropy} to obtain fully consistent results for dynamical observables like the mean-squared displacement~\cite{breoni2020active, feng2022mode}, which reveals a mass-independent long-time diffusive behavior of the single particle.
	Moreover, these theoretical models have been employed to evaluate the effect of inertia on the collective phenomena typical of active matter.
	It was found that (translational) inertia reduces MIPS~\cite{mandal2019motility, dai2020phase, su2021inertia, omar2021tuning}, hinders the crystallization~\cite{de2020phase, liao2021inertial}, promotes hexatic ordering~\cite{negro2022inertial} in homogeneous phases and, in general, reduces the spatial velocity correlations characterizing dense active systems~\cite{caprini2021spatial, caprini2021collective, marconi2021hydrodynamics} both the liquid and solid state.

	The second, and arguably the more critical, step is to include the effect of a non-vanishing moment of inertia affecting the rotational motion.
	This ingredient is fundamentally relevant in granular experiments to reproduce the inertial delay, i.e., the temporal delay between the active force and particle velocity observed for a single active granular particle~\cite{scholz2018inertial}.
	To model such a generic inertial active particle not only the overdamped translational equation of motion but also the stochastic process describing the dynamics of the active velocity (or active force) itself needs to be modified.
	Again, this can be quite naturally achieved by a second extension of the ABP dynamics through including inertia on the rotational velocity~\cite{scholz2018inertial, crosato2019irreversibility, gutierrez2020inertial, herrera2021maxwell, sprenger2021time, hecht2022active}.
	Using this inertial ABP (or active Langevin) model, it has been found that the long-time dynamics are strongly affected by a nonzero moment of inertia.
	Successively, the effect of rotational inertia in systems of interacting particles has been investigated and identified as a strategy to promote collective phenomena~\cite{caprini2022role, de2022collective}. 
    Finally, rotational inertia has been recently considered also in macroscopic descriptions, such as active phase crystal model~\cite{te2021jerky, arold2020time}, to investigate sound waves in active matter.
	
	Despite the success of AOUPs for describing overdamped active particles or active particles with translational inertia, a comprehensive inertial AOUP model, i.e., a Gaussian process for the active velocity also accounting for rotational inertia, has not been properly introduced.
	While such an achievement would be helpful in view of making further theoretical progress, this challenge is complicated by the intrinsic coupling between the angular dynamics and those of the modulus of the active velocity~\cite{caprini2022parental}, preventing conformance with inertial ABPs.
	A first attempt to do so has been introduced in Ref.~\cite{lisin2022motion} by mapping rotational inertia onto effective parameters of the AOUP model.

\begin{figure}
		\includegraphics[width=\columnwidth]{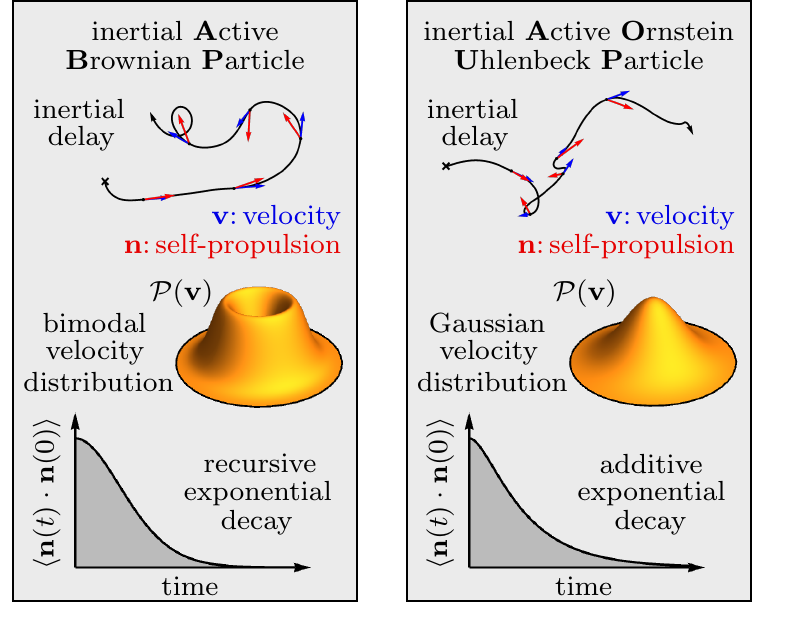}
		\caption{Schematic comparison of two models for active particles displaying both translational and rotational inertia. The left panel shows an inertial active Brownian particle (ABP) \cite{sprenger2021time}, while the right panel shows an inertial active Ornstein-Uhlenbeck particle (AOUP), introduced here through Eqs.~\eqref{eq:langevin_trans} and~\eqref{eq:inertial_AOUP}. 
		\textbf{Top:} both models display persistent trajectories with inertial delay (the particle velocity $\vec{v}(t)$ lags behind the self-propulsion vector $\vec{n}(t)$).
		\textbf{Middle:} the overall velocity distribution $\mathcal{P}(\vec{v})$ of the inertial AOUP has the advantageous Gaussian form, while that of the inertial ABP is bimodal.
		\textbf{Bottom:} The autocorrelation functions $\mean{\vec{n}(t) \cdot \vec{n}(0)}$ of the self-propulsion vector have a different form, contrast the recursive exponential decay  (Eq.~\eqref{eq:orientational_correlation}) with the additive exponential decay (Eq.~\eqref{eq:correlation_inertial_AOUP}), but each model incorporates three characteristic time scales of inertial active motion, see Eq.~\eqref{eq:ABPparsTAU} or also Eq.~\eqref{eq:AOUPparsTAU}.
		}
		\label{fig:inertial_abp_vs_aoup}
	\end{figure}

	In this paper, we propose a generalization of the inertial active Ornstein-Uhlenbeck particle (AOUP) model incorporating the characteristic time scales of active particles with both translational and rotational inertia.
	As illustrated in Fig.~\ref{fig:inertial_abp_vs_aoup}, this ensures that, in analogy to the inertial ABP, the decay of the autocorrelation function of the self-propulsion vector takes longer that the single-exponential decay for zero moment of inertia \cite{scholz2018inertial}. 
	As a result, both models consistently predict persistent trajectories, which also show inertial delay~\cite{scholz2018inertial,sprenger2021time}.
	However, the velocity distribution of the inertial AOUP has, by construction, a Gaussian shape at variance with the bimodal shape of the inertial ABP~\cite{herrera2021maxwell}.

	The paper is structured as follows.
	We first provide in Sec.~\ref{sec_ABP} a rundown of the inertial ABP model and discuss briefly the effect of rotational inertia on the persistence time of the active motion.
	Then, in Sec.~\ref{sec_AOUP}, we extend the inertial AOUP to account for rotational inertia. 
	Subsequently, in Sec.~\ref{sec_RES} we discuss the dynamical predictions for the time-dependent orientational correlation, velocity autocorrelation, delay function, as well as the mean and mean-square displacement.
    To validate the inertial AOUP model introduced in this paper, we compare the results for appropriately identified parameters to those of the inertial ABP. 
	Finally, we present a conclusive discussion in Sec.~\ref{sec_CON}.

	\section{Inertial ABP model \label{sec_ABP}}
	We consider an inertial self-propelled particle in two spatial dimensions, characterized by its mass $m$ and moment of inertia $J$. 
	The particle dynamics is described by stochastic evolution for the center-of-mass velocity $\vec{v} = \dot{\vec{r}}$ (with $\vec{r}$ being the center-of-mass position) and the angular velocity $\omega = \dot{\phi}$ (with $\phi$ being the orientational angle of the particle). 
	The translational motion is governed by Newton's second law of motion
	\begin{subequations}\label{eq:langevin_trans}
		\begin{align}
		\dot{\vec{r}} &= \vec{v}\,, \label{eq:langevin_r} \\
		m \, \dot{\vec{v}} &= -  \gamma \, \vec{v} - \Nabla U(\vec{r}) + \gamma \sqrt{2 D_t} \vec{\xi} +\gamma v_0 \vec{n}\,,
		\label{eq:langevin_v}
		\end{align}
	\end{subequations}
	where the acceleration term $m \, \dot{\vec{v}}$ accounts for translational inertia.
	The total force on the right-hand-side of Eq.~\eqref{eq:langevin_v} is given by the sum of the friction force $-\gamma \vec{v}$, proportional to the translational frictions coefficient $\gamma$, the external force $-\Nabla U(\vec{r})$ with the potential $U(\vec{r})$, and the thermal force $\gamma \sqrt{2 D_t} \vec{\xi}$, whose intensity is given by the translational diffusion coefficient $D_t$ and distributed like a zero-mean unit variance Gaussian white noise $\vec{\xi}$.
	Finally, the active force $\gamma v_0 \vec{n}$ couples via the orientation vector, $\vec{n} = (\cos \phi, \sin \phi)$, the translational motion to a rotational degree of freedom. 
	In the ABP model the modulus of the active force is constant and sets the self-propulsion speed $v_0$.
	
	In a similar manner, the rotational motion
	\begin{subequations}
		\label{eq:langevin_rot}
		\begin{align}
		\dot{\phi} &= \omega\,, \label{eq:langevin_phi}\\
		J \, \dot{\omega} + \gamma_\text{r} \, \omega &= \gamma_\text{r} \sqrt{2 D_\text{r}} \eta\,, \label{eq:langevin_omega}
		\end{align}
	\end{subequations} 
	involves a friction torque $-\gamma_\text{r} \omega$ with the rotational friction coefficient $\gamma_\text{r}$ and a stochastic torque $\gamma_\text{r} \sqrt{2 D_\text{r}} \eta$, where the effective rotational diffusion coefficient $D_\text{r}$ quantifies the rotational noise strength and the Gaussian noise $\eta$ has zero-mean and unit variance. Here, the angular acceleration term $J \, \dot{\omega}$ accounts for rotational inertia.
	Overall, the inertial ABP is characterized by three typical times
		\begin{equation}\label{eq:ABPparsTAU} 
	\tau:=\frac{1}{D_\text{r}}\,,\ \ \ \tau_J:= \frac{J}{\gamma_\text{r}}\,,\ \ \ \tau_m:=\frac{m}{\gamma}\,,
	\end{equation}
	representing rotational diffusion, translational memory and rotational memory, respectively. In what follows, we use $\tau$ as the unit time.

	By taking a closer look at Eq.~\eqref{eq:langevin_omega}, we see that the angular velocity $\omega$ is described by an Ornstein-Uhlenbeck process such that 
	\begin{equation}
	  \mean{ \omega(t)\, \omega(0)} = \frac{1}{\tau\tau_J} e^{-t/\tau_J} \,.
	    \label{eq:omega_OUP}
	\end{equation}
	Thus, the time scale $\tau_J$, entering in Eq.~\eqref{eq:omega_OUP}, introduces memory in the angular velocity, such that the orientational correlation function
	\begin{align}  \label{eq:orientational_correlation}
	\mean{\vec{n}(t) \cdot \vec{n}(0)} = e^{- (t/\tau-\tau_J/\tau (1-e^{-t / \tau_J}))}
	\end{align}
	exhibits a \textit{recursive} exponential decay (see footnote~\cite{footnoteDECAY} for a clarification of the use of the term ``recursive''), instead of the single-exponential decay in the absence of rotational inertia, $\tau_J \to 0$.
	In particular, this orientational correlation decays quadratically for short times, 
	\begin{equation}  \label{eq:orientational_correlation_short_time_abp}
	\mean{\vec{n}(t) \cdot \vec{n}(0)} = 1 - t^2 / (2 \tau\tau_J) + \O{t^3},
	\end{equation}
	and the overdamped result $\tau_\text{p} = \tau$ for the characteristic persistence time $\tau_\text{p} = \int_{0}^{\infty} \! \mean{\vec{n}(t) \cdot \vec{n}(0)} \mathrm{d}t$ generalizes to
	\begin{equation} \label{eq:persistence_time}
	\tau_\text{p} = \tau_J \, e^{\tau_J/\tau} \, (\tau_J/\tau)^{-\tau_J/\tau} \, \Gamma(\tau_J/\tau,0,\tau_J/\tau)\,,
	\end{equation}
	where $\Gamma (x,z_0,z_1) = \integral{z_0}{z_1}{t^{x-1} e^{-t}}{t}$ is the incomplete gamma function.
	
	In general, the persistence time $\tau_\text{p}$ increases when $\tau$ is increased. Compared to the overdamped case, this increase is more significant when the typical time $\tau_J$ (or the moment of inertia) is increased.
	Therefore, it is apparent that inertial effects hinder the particle's ability of changing the direction of its self-propulsion vector in response to an applied torque. 
Further results for an inertial ABP are contained in Appendix~\ref{app_ABP}.
It should be noted that, due to the implicit dependence of most quantities on $\tau_J$, such as $\tau_\text{p}$ in Eq.~\eqref{eq:persistence_time}, a comprehensive analytical picture is impaired.
Explicit analytical insight can be obtained in the small-rotational-inertia limit.

	\subsection{ABP for small rotational inertia}

	Neglecting rotational inertia, $\tau_J \to 0$, the rotational dynamics of the inertial ABP coincides with the usual ones, expected for overdamped ABP.
	In this limit, the angular velocity $\omega$ converges onto a zero-mean $\delta$-correlated Gaussian white noise with 
	\begin{equation}
	    \mean{ \omega(t) \omega(t')} \sim \frac{2}{\tau} \delta(t-t') 
	\end{equation}
	as a result of the asymptotic limit of Eq.~\eqref{eq:omega_OUP}.
	Secondly, expanding Eq.~\eqref{eq:orientational_correlation} in powers of $\tau_J$, the autocorrelation of the orientational vector, $\mathbf{n}$, reads
		\begin{subequations}
		\label{eq:orientational_correlation_od}
		\begin{align}
		&\mean{\vec{n}(t)\cdot \vec{n}(0)} = e^{- t/\tau} \Big( 1 + \tau_J/\tau + \O{\tau_J^2} \Big) \label{eq:orientational_correlation_od_LIN}\\
		&=e^{- t/\tau} \Big( 1 + \tau_J/\tau(1-e^{-t / \tau_J}) + \O{\tau_J^2} \Big)\,, \label{eq:orientational_correlation_od_EXP}
		\end{align}
	\end{subequations} 
	where the second equality conveniently retains $\tau_J$ as a typical exponential decay time. 
	From Eq.~\eqref{eq:orientational_correlation_od}, we can naturally identify the persistence time, $\tau_\text{p}$, as the inverse of the rotational diffusion coefficient, $\tau$, in the overdamped limit $\tau_J \to 0$.
	This can be explicitly verified by expanding Eq.~\eqref{eq:persistence_time} in powers of $\tau_J$, such that
	\begin{equation}
	   \tau_\text{p}= \tau + \tau_J - \frac{\tau_J^2}{2\tau} + \O{\tau_J^3} \,,
	    \label{eq:taup_Taylor}
	\end{equation}
	and then considering the limit $\tau_J \to 0$.

	\section{Fully inertial AOUP model \label{sec_AOUP}}
	Despite the simplicity and intuitive nature of the ABP model, obtaining analytical results that go beyond the potential-free particle is not an easy task \cite{caraglio2022analytic}, even more so, in the presence of rotational inertia.
	The AOUP model, initially proposed in overdamped systems (without inertia) represents an alternative and simplified model to the ABP which is obtained by replacing the orientation vector $\vec{n}$ in Eq.~\eqref{eq:langevin_v} by an Ornstein-Uhlenbeck process with correlation time $\tau$ and unit variance.
	This simple approach works well because the autocorrelation $\mean{\vec{n}(t) \cdot \vec{n}(0)}$ of both an (overdamped) ABP and AOUP has the same exponential shape decaying with a typical correlation time, that coincides with the common persistence time. 
	To get consistent results, it is merely required to ensure that $\tau\equiv 1/D_\text{r}$ describing the AOUP dynamics represents the inverse rotational diffusion coefficient of the ABP in two dimensions~\cite{farage2015effective}, as we imply here through Eq.~\eqref{eq:ABPparsTAU}. 
	In the AOUP case, the whole stochastic process modeling the active self-propulsion vector $\vec{n}$ is Gaussian, thus offering a simplified platform to derive analytical results in the presence of interactions and external potentials~\cite{caprini2022parental}.

	The inclusion of translational inertia in the AOUP model has been proposed and investigated \cite{caprini2020inertial, nguyen2021active}, and, in general, does not present any additional conceptual or technical difficulties compared to the ABP model.
	This is because translational inertia does not affect the dynamics of the active force, i.e., the orientational angle $\phi$ in the ABP case or the Ornstein-Uhlenbeck process for the self-propulsion vector $\vec{n}$ (see below) in the AOUP case. 
	In the presence of rotational inertia, pursuing a similar strategy of deriving a Gaussian approximation to the ABP dynamics is not straightforward because of the intricate structure of Eq.~\eqref{eq:orientational_correlation}, which does no longer posses a single-exponential shape as in the overdamped case, Eq.~\eqref{eq:orientational_correlation_od_LIN}.
	Intuitively, a minimal description of rotational inertia requires (i) an additional time scale, $\tau_J$, and (ii) an additional scaling factor, $\tau_J/\tau$, both related to the moment of inertia, which affects the angular velocity autocorrelation.

	To generalize the AOUP model to the presence of rotational inertia, we introduce an additional colored noise $\vec{\chi}$ in the dynamics of the self-propulsion vector $\mathbf{n}$, characterized by its own rotational memory time $\tau_\chi$ and the noise strength $D_\chi/\tau_\chi^2$, so that $\mathbf{n}$ evolves as
	\begin{subequations}
		\label{eq:inertial_AOUP}
		\begin{gather} 
		\dot{\vec{n}} = -\frac{\vec{n}}{\tau} + \sqrt{\frac{1}{\tau}} \vec{\chi}, \label{eq:langevin_n} \\
		\dot{\vec{\chi}} = -\frac{\vec{\chi}}{\tau_\chi} + \frac{\sqrt{2 D_\chi}}{\tau_\chi} \vec{\zeta} \,. \label{eq:langevin_chi}
		\end{gather}
	\end{subequations}
	This model ensures that	Eq.~\eqref{eq:langevin_n} formally coincides with the overdamped AOUP model, i.e., when the auxiliary process $\boldsymbol{\chi}$ is a white noise. 
	Here, the additional Ornstein-Uhlembeck process for $\boldsymbol{\chi}$, evolving according to Eq.~\eqref{eq:langevin_chi}, prescribes a more general colored noise.
	As a consequence, the rotational AOUP model is not only characterized by one typical time $\tau$ (which in overdamped systems coincides with the persistence time), but also by an additional time $\tau_\chi$ and the inertial diffusivity $D_\chi$.
	The latter can be conveniently determined as
		\begin{align}
 D_\chi = \frac{\tau+ \tau_\chi}{2\tau}  
	\end{align}
	from the condition
		\begin{equation}
		\label{eq:conditions1}
		\mean{\vec{n}(0) \cdot \vec{n}(0)} =  1\,, \\
	\end{equation}
    ensuring the unitary normalization of $\vec{n}(t)$ (which is a unit vector in the ABP case) to set the velocity scale by $v_0$ without ambiguity~\cite{caprini2022parental}.

	The standard AOUP model in the overdamped limit is naturally achieved by requiring $\tau_\chi \to 0$ and $D_\chi\to1/2$ such that Eq.~\eqref{eq:langevin_chi} reduces to a white noise with zero average and unit variance.
	Moreover, the linearity of Eq.~\eqref{eq:inertial_AOUP} allows us to analytically derive the autocorrelation function
	\begin{equation}
	\mean{\vec{n}(t) \cdot \vec{n}(0)} = \frac{2 D_\chi \tau}{\tau^2 - \tau_\chi^2} \left( \tau \, e^{-t/\tau} - \tau_\chi \, e^{-t/\tau_\chi} \right) \,
	\label{eq:correlation_inertial_AOUP}
	\end{equation}
  of the self-propulsion vector $\mathbf{n}$,
	which is characterized by an \textit{additive} exponential decay (see footnote~\cite{footnoteDECAY} for a clarification of the use of the term ``additive''),
	i.e., the superposition of two exponential functions with the correlation times $\tau$ and $\tau_\chi$.
	Comparing this result to the expansion in Eq.~\eqref{eq:orientational_correlation_od_EXP} for the inertial ABP, we deduce that the structure of Eq.~\eqref{eq:correlation_inertial_AOUP} with two different decay times constitutes the minimal ingredient to account for rotational inertia.
    In the rest of this work, we validate the inertial AOUP model by establishing a suitable relation between the parameters $\tau_\chi$ and $D_\chi$ in and those, $\tau$ and $\tau_J$, of the inertial ABP model to quantify the impact of rotational inertia through Eq.~\eqref{eq:inertial_AOUP}.

	\subsection{Relation to inertial ABP model}

	Comparing the full predictions of the two models for the autocorrelation function given by Eq.~\eqref{eq:orientational_correlation} and Eq.~\eqref{eq:correlation_inertial_AOUP}, it becomes apparent that, at variance with the overdamped limit ($\tau_\chi\to0$ or $\tau_J\to0$), the shape of $\mean{\vec{n}(t) \cdot \vec{n}(0)}$ does not coincide, see also Fig.~\ref{fig:inertial_abp_vs_aoup}.
	To provide a coherent scheme for identifying the rotational memory time $\tau_\chi$ of our inertial AOUP model, we impose here, in addition to Eq.~\eqref{eq:conditions1}, the second  natural condition
	\begin{equation}
		\label{eq:conditions2}
		\int_{0}^{\infty} \! \mean{\vec{n}(t) \cdot \vec{n}(0)}\, \mathrm{d}t = \tau_\text{p}\,
	\end{equation}
	satisfied by the inertial ABP model, which enforces that both models have the same autocorrelation time and, thus, predict the same long-time diffusion behavior.
	Thus we can identify the parameters of both models according to
 \begin{subequations}
 \label{eq:AOUPpars}
	\begin{align}
	\tau_\chi &= \tau_\text{p} - \tau=\tau_J+\O{\tau_J^2}  \label{eq:AOUPparsTAU} \,,\\ 
 D_\chi &= \frac{\tau_\text{p}}{2 \tau}=\frac{1}{2}+\frac{\tau_J}{2\tau}+\O{\tau_J^2}  \,,
	\end{align}
 	\end{subequations}
	where the provided small-$\tau_J$ expansions are apparent from Eq.~\eqref{eq:taup_Taylor}.

\begin{figure}
		\includegraphics[width=0.9\columnwidth]{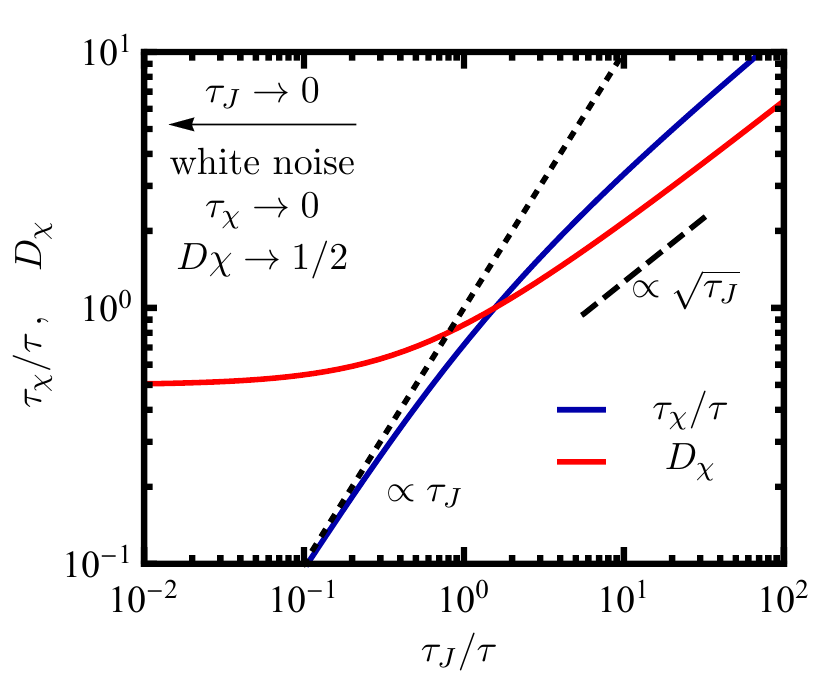}
		\caption{Additional parameters of the inertial AOUP model, $\tau_\chi/\tau$ (blue curve) and $D_\chi$ (red curve), related to the inertial ABP model through Eq.~\eqref{eq:AOUPpars} as a function of the normalized rotational memory time $\tau_J$ of the inertial ABP. Dashed black lines indicate the scaling $\sim\sqrt{\tau_J}$ occurring for large $\tau_J$ while for $\tau_J\to0$ the limiting values $D_\chi\to1/2$ and $\tau_\chi\to0$ are approached and $\tau_\chi$ scales as $\sim\tau_J$ (dotted black line).
		}
		\label{fig:inertial_parameters}
	\end{figure}

	 To understand the meaning of the new inertial parameters $\tau_\chi$ and $D_\chi$ of our generalized AOUP model, their values are shown in Fig.~\ref{fig:inertial_parameters} as a function of the rotational memory time $\tau_J$ of the inertial ABP.  
	It can be seen that both parameters $\tau_\chi$ and $D_\chi$ are increasing functions of $\tau_J$. They scale as $\sim\!\sqrt{\tau_J/\tau}$ for $\tau_J/\tau\gg1$, as can be deduced from the function $\tau_\text{p}$ given by Eq.~\eqref{eq:persistence_time}.
	 Moreover, the limit of vanishing rotational inertia, $\tau_J\to0$, is consistent with the overdamped AOUP, since according to Eq.~\eqref{eq:taup_Taylor} the persistence time $\tau_\text{p}$ reduces to $\tau$, such that we observe the limits $\tau_\chi\to 0$ and $D_\chi \to 1/2$. 
 As a consequence, Eq.~\eqref{eq:langevin_n} reduces in the overdamped limit to a standard Ornstein-Uhlenbeck process with the white noise $\vec{\chi}=\vec{\zeta}$, employed to describe active particles without rotational inertia.
 Further aspects of the small-rotational-inertia limit are discussed in Sec.~\ref{sec_AOUPsI}.
 In the opposite limit, $\tau_J\to\infty$, the inertial AOUP persistently moves along a straight line, which is consistent with the inertial ABP.
Thus our model accurately includes both limits of vanishing and infinite rotational inertia.
	
\subsection{Probability densities for inertial AOUPs \label{sec:Pvn}}
	
	One of the main advantages of the inertial AOUP model, defined by Eqs.~\eqref{eq:langevin_trans} and~\eqref{eq:inertial_AOUP}, is that the stationary probability density $\fct{P}(\vec{v}, \vec{n}, \vec{\chi})$ can be explicitly derived via its correlation matrix. 
	We list these results in Appendix~\ref{aoup:steady_states}.
	Here, we discuss the reduced probability $\fct{P}(\vec{v}, \vec{n})$ to find a given velocity $\vec{v}$ and self-propulsion $\vec{n}$ which is obtained via integration of the full probability density with respect to the auxiliary process $\vec{\chi}$. 
	The distribution $\fct{P}(\vec{v}, \vec{n})$ can be expressed as
    \begin{equation}
		\fct{P}(\vec{v}, \vec{n}) = \fct{P}(\vec{v}\vert\vec{n}) \fct{P}(\vec{n}) , \label{eq:probability_v_n}
	\end{equation}
	where $\fct{P}(\vec{n})$ is the marginal probability density of the self-propulsion vector $\vec{n}$ with unit-variance, thus
	\begin{equation}
		\fct{P}(\vec{n}) \propto \exp{ \big( - \vec{n}^2 \big)},
	\end{equation}
	and $\fct{P}(\vec{v}\vert\vec{n})$ defines the conditional probability to find a particle at a velocity $\vec{v}$ with prescribed $\vec{n}$. 
	Using the time scales $\tau$ and $\tau_m$ from Eq.~\eqref{eq:ABPparsTAU} and the rotational memory time $\tau_\chi$ of the inertial AOUP, we have
	\begin{subequations}\label{eq:Pvn}
		\begin{align}
			\fct{P}(\vec{v}\vert\vec{n}) \propto & \exp \bigg( - \frac{\big( \vec{v} - \mean{\vec{v}\vert\vec{n}} \big)^2}{ \sigma(\vec{v} \vert \vec{n}) }   \bigg) , \\  
			\mean{\vec{v} \vert \vec{n}} = & \frac{v_0}{\tau - \tau_\chi} \bigg( \frac{\tau^2}{\tau + \tau_m} - \frac{\tau_\chi^2 }{\tau_\chi + \tau_m} \bigg)  \, \vec{n}, \label{eq:condvn} \\
			\sigma(\vec{v} \vert \vec{n}) = &  \frac{2 D_t}{\tau_m} + \frac{v_0^2 \tau_m^2 ( \tau \tau_m  + \tau_m \tau_\chi + \tau \tau_\chi)}{(\tau + \tau_m)^2 (\tau_\chi+\tau_m)^2},
		\end{align}
	\end{subequations}
	where $\fct{P}(\vec{v}\vert\vec{n})$ is centered around the conditional average $\mean{\vec{v} \vert \vec{n}}$ of $\vec{v}$ at given $\vec{n}$ with its corresponding conditional variance $\sigma(\vec{v} \vert \vec{n})$.

	 By integrating the distribution $\mathcal{P}(\mathbf{v}|\mathbf{n})$ in Eq.~\eqref{eq:probability_v_n} over $\mathbf{n}$, we derive the velocity distribution of a system of ideal inertial AOUPs
	\begin{subequations}
		\begin{align}
			\fct{P}(\vec{v}) &\propto \exp{\left( - \frac{\vec{v}^2}{ \mean{\vec{v}^2}} \right)}, \\  
			\mean{\vec{v}^2} & =  \frac{2 D_t}{ \tau_m } + \frac{v_0^2}{\tau - \tau_\chi} \bigg( \frac{\tau^2}{\tau + \tau_m} - \frac{\tau_\chi^2 }{\tau_\chi + \tau_m} \bigg)   , \label{eq:msv}
		\end{align}
	\end{subequations}
    with the mean-square velocity $\mean{\vec{v}}^2$. 
    Such a distribution has a typical Boltzmann-like shape as illustrated in Fig.~\ref{fig:inertial_abp_vs_aoup}, with an effective temperature determined by the swim velocity $v_0$ and the three typical time scales $\tau$, $\tau_m$, and $\tau_\chi$.

	\subsection{AOUP for small rotational inertia \label{sec_AOUPsI}}
	
	In the absence of rotational inertia, the inertial AOUP model converges onto the standard AOUP model employed to describe overdamped active particles or active particles with translational inertia only.
This is evident by taking the overdamped limit in Eq.~\eqref{eq:langevin_chi}, i.e., considering $\tau_\chi\to0$.
The nature of the Ornstein-Uhlenbeck process $\boldsymbol{\chi}$ allows us to derive the steady-state autocorrelation 
	\begin{equation}
	   \mean{\vec{\chi}(t) \cdot \vec{\chi}(0)} = 2\frac{D_\chi}{\tau_\chi} e^{-t/\tau_\chi}
    = \frac{\tau+\tau_J}{\tau\tau_J} e^{-t/\tau_J}+\O{\tau_J^2}\,,
	    \label{eq:CHI_OUP}
	\end{equation}
 where, in the last equality, we have used Eqs.~\eqref{eq:AOUPpars} holding at first order in $\tau_J$.
  This shape links the correlator of $\vec{\chi}$ to the correlator, Eq.~\eqref{eq:omega_OUP}, of the angular velocity $\omega$ in the inertial ABP model.
  This confirms our identification of the additional degree of freedom $\boldsymbol{\chi}$ in the inertial AOUP model as the key dynamical variable able to capture the effects of rotational inertia.
  To see this, we further note that, in Cartesian coordinates, the dynamics of the self-propulsion vector $\mathbf{n}$ can be expressed as $\dot{\mathbf{n}}=\mathbf{n}\times \mathbf{z}\,\omega$, where $\mathbf{z}$ is the unit vector perpendicular to the two-dimensional plane of motion~\cite{sprenger2021time}. 
  Similarly to the overdamped case~\cite{caprini2022parental}, the AOUP approximation can then be imagined as replacing this term by an Orstein-Uhlenbeck process.

 In the same spirit, the autocorrelation~\eqref{eq:correlation_inertial_AOUP} of the self-propulsion vector $\vec{n}$ can be expanded with the help of Eq.~\eqref{eq:AOUPpars} as
	\begin{equation}
	    \mean{\vec{n}(t) \cdot \vec{n}(0)} = \frac{1}{\tau - \tau_J} \left( \tau \, e^{-t/\tau} - \tau_J \, e^{-t/\tau_J} \right) +\O{\tau_J^2}\,.
	\end{equation}
	By additionally setting $\tau\gg\tau_J$ for small moment of inertia, we recover Eq.~\eqref{eq:orientational_correlation_od_EXP}.
 Thus, our model goes beyond a naive mapping of this small-rotational-inertia limit.

    In addition, the probability distribution $\fct{P}(\vec{v}\vert\vec{n})$ in Eq.~\eqref{eq:Pvn} converges to the one found in Ref.~\cite{caprini2020inertial} without rotational inertia.
    By expanding for small $\tau_J$, mean and variance of the Gaussian distribution reads 
	\begin{subequations}\label{eq:PvnOD}
		\begin{align}
			\mean{\vec{v} \vert \vec{n}} = & \frac{v_0 \tau}{\tau + \tau_m} \, \vec{n} +  \frac{v_0 \tau_J}{\tau + \tau_m} \, \vec{n} +\O{\tau_J^2}, \\
			\sigma(\vec{v} \vert \vec{n}) = &  \frac{2 D_t}{\tau_m} + \frac{v_0^2 \tau_m \tau}{(\tau + \tau_m)^2} - \frac{v_0^2 (\tau - \tau_m )\tau_J}{(\tau + \tau_m)^2}+\O{\tau_J^2},
		\end{align}
	\end{subequations} 
    where we have neglected order $\tau_J^2$.
    The zero-order result in Eq.~\eqref{eq:PvnOD} coincides with the variance calculated in Ref.~\cite{caprini2020inertial}, while the first correction in $\tau_J$ decreases the velocity variance if $\tau>\tau_J$ (long-persistent regime) and increases the variance in the opposite limit.

	\section{Comparison between inertial ABP and inertial AOUP} \label{sec_RES}
	The inertial AOUP model introduced in Sec.~\ref{sec_AOUP} defines a purely Gaussian process which  in general significantly simplifies the theoretical analysis compared to the inertial ABP model, specified in Sec.~\ref{sec_ABP}. 
	However, at variance with the overdamped case, there is no one-to-one identification of the parameters in these two models, since the shape of the autocorrelations, Eqs.~\eqref{eq:orientational_correlation} and~\eqref{eq:correlation_inertial_AOUP}, does not coincide. 
	Therefore, a careful comparison between the inertial ABP and inertial AOUP is needed. 
	To this end, we evaluate in the following several observables for different values of the rotational memory time $\tau_J$ of the inertial ABP
 which sets the corresponding rotational memory time $\tau_\chi$ of the inertial AOUP through Eqs.~\eqref{eq:AOUPparsTAU} and \eqref{eq:persistence_time}.
We thus	explore all regimes where the rotational inertia plays a marginal ($\tau_J \ll \tau$), intermediate ($\tau_J \approx \tau$) and relevant ($\tau_J \gg \tau$) role. 
 In particular, we compare the autocorrelation of the self-propulsion vector, velocity correlations and the cross correlation between self-propulsion vector and velocity, known as delay function.
	Finally, we consider the mean-square displacement and the long-time diffusion coefficient.
	The implicit analytic reference results for the inertial ABP model are listed in Appendix~\ref{app_ABP}.

	\subsection{Orientational correlation function \label{sec:ocf}}
	
	\begin{figure}
		\includegraphics[width=0.9\columnwidth]{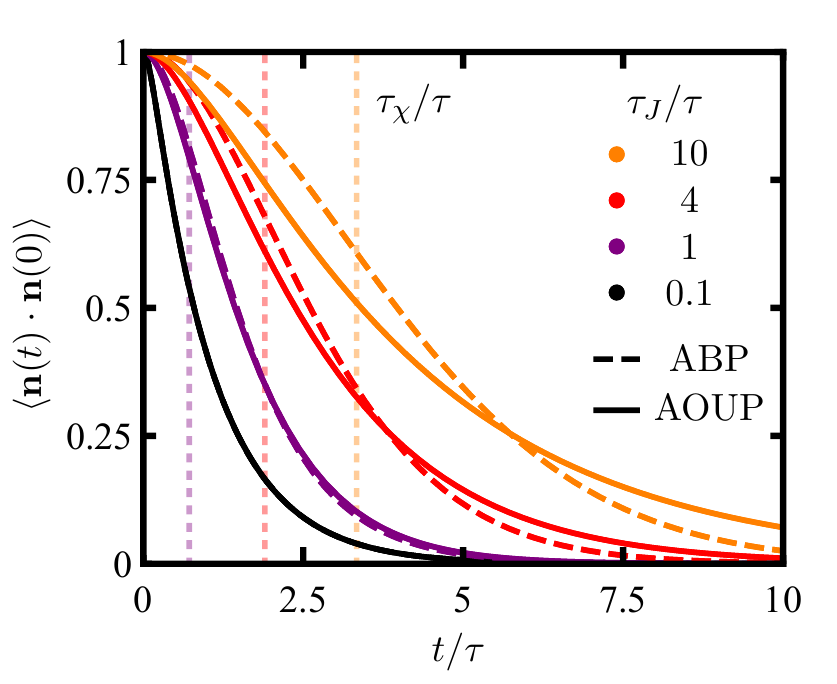}
		\caption{Orientational correlation $\mean{\vec{n}(t) \cdot \vec{n}(0)}$ as a function of time $t/\tau$ for different moments of inertia given through $\tau_J/\tau$ (as labeled). Solid and dashed lines correspond to an AOUP and ABP, respectively.
		Vertical dotted lines indicate the rotational memory time $\tau_\chi$ of the inertial AOUP.} 
		\label{fig:orientation_correlation}
	\end{figure}

 Having established in Sec.~\ref{sec_AOUPsI} that both inertial ABP and AOUP models yield the same autocorrelation function $\mean{\vec{n}(t) \cdot \vec{n}(0)}$ of the self-propulsion vector $\vec{n}$ for small rotational inertia, we provide in Fig.~\ref{fig:orientation_correlation} a comparison for different reduced moments of inertia $\tau_J/\tau$. 
 As expected, for $\tau_J \ll \tau$ and $\tau_J \approx \tau$, a good agreement is obtained on all timescales (see the comparison between solid and dashed lines).
 However, for $\tau_J \gg \tau$, we observe small deviations between the two models. 
 In particular, the inertial AOUP model predicts a faster early decay, which we understand from comparing the short-time expansion
 	\begin{equation}
	\mean{\vec{n}(t) \cdot \vec{n}(0)} = 1 - \frac{t^2}{2 \, \tau \tau_\chi} +\O{t^3} 
	\end{equation}
 to the ABP result in Eq.~\eqref{eq:orientational_correlation_short_time_abp} and recognizing that $\tau_\chi \leq \tau_J$ (see Fig.~\ref{fig:inertial_parameters}). 
 At later times, there is a crossover between $\tau_\chi\lesssim t\lesssim\tau_J$ as the autocorrelation of the inertial AOUP has a longer decay tail,
 which reflects the nature of the additive exponential decay, compared to the faster recursive exponential decay the of inertial ABP.

	\subsection{Velocity correlation function}

	\begin{figure}
		\includegraphics[width=0.9\columnwidth]{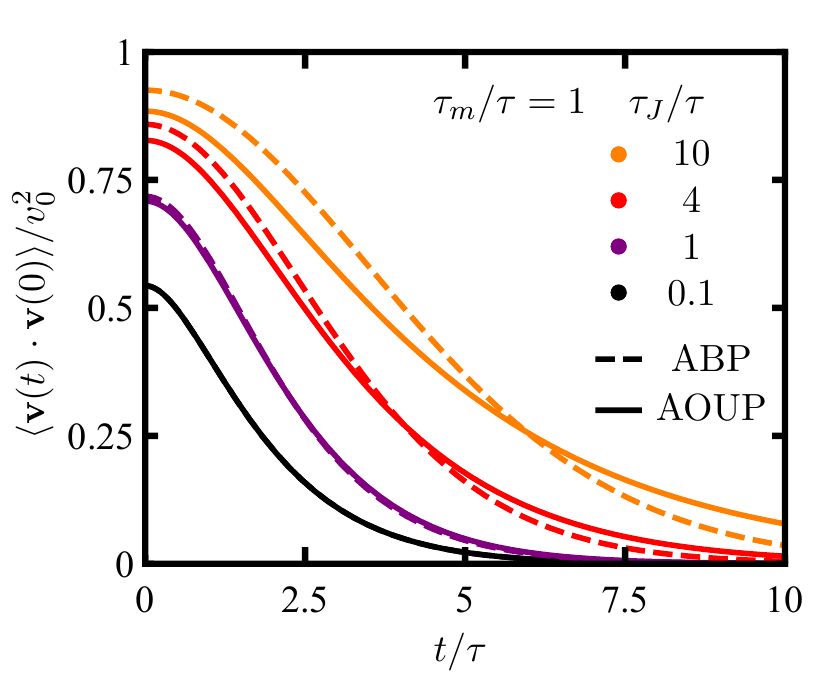}
		\caption{Velocity correlation function, $\mean{\vec{v}(t) \cdot \vec{v}(0)}$, as a function of time $t/\tau$ for a fixed mass $m$ given through $\tau_m/\tau=1$ different moments of inertia $J$ given through $\tau_J/\tau$ (as labeled). Solid and dashed lines correspond to an AOUP and ABP, respectively.}
		\label{fig:velocity_correlation}
	\end{figure}

 The velocity correlation of the inertial AOUP reads
	\begin{equation} \label{eq:velocitycorrelation}
	\mean{\vec{v}(t) \cdot \vec{v}(0)} = \frac{2 D_t}{ \tau_m } e^{-t/\tau_m} + \frac{v_0}{2} \Big( \mean{\vec{v}(t) \cdot \vec{n}(0)} + \mean{\vec{v}(0) \cdot \vec{n}(t)}  \Big),
	\end{equation}
	with	
	\begin{subequations}\label{eq:vini}
		\begin{align}
	\mean{ \vec{v}(t) \cdot \vec{n}(0)} = & \frac{v_0 }{\tau - \tau_\chi} \bigg( \frac{\tau^2 e^{-t/\tau} }{\tau - \tau_m}  - \frac{\tau_\chi^2 e^{-t/\tau_\chi}}{\tau_\chi - \tau_m}   \bigg)   \nonumber \\
	& + \frac{2 v_0 \tau_m^3 (\tau + \tau_\chi) e^{-t/\tau_m}}{(\tau^2 - \tau_m^2)(\tau_\chi^2 - \tau_m^2)}  , \\
	\mean{ \vec{v}(0) \cdot \vec{n}(t)} = & \frac{v_0 }{\tau - \tau_\chi} \bigg( \frac{\tau^2 e^{-t/\tau} }{\tau + \tau_m} - \frac{\tau_\chi^2 e^{-t/\tau_\chi} }{\tau_\chi + \tau_m}   \bigg).
	\end{align}
	\end{subequations}
 This result serves as a closed-form approximation for the inertial ABP result.
 As shown in Fig.~\ref{fig:velocity_correlation}, our model consistently predicts stronger velocity correlations for all times when the rotational inertia is increased.
 The small deviations for large moment of inertia and the crossover of the decay behavior are quite similar to those discussed in Sec.~\ref{sec:ocf} for the orientational correlation function.

In addition, we observe in Fig.~\ref{fig:velocity_correlation} an offset at $t=0$ between the two models, i.e., they predict a distinct mean-square velocity $\mean{\vec{v}^2}\equiv\mean{\vec{v}(t) \cdot \vec{v}(0)}$.
For the inertial AOUP, we recover the result for $\mean{\vec{v}^2}$ given by Eq.~\eqref{eq:msv}.
We can thus conclude that rotational inertia increases the translational kinetic temperature (which is proportional to $\mean{\vec{v}^2}$). 
Taking the limit $\tau_J\to\infty$ in Eq.~\eqref{eq:msv}, we find that $\mean{\vec{v}^2}\to 2 D_t / \tau_m + v_0^2$
reaches a plateau value which is the same as found for an inertial ABP.

	\subsection{Delay Function}
	
	\begin{figure}
		\includegraphics[width=0.9\columnwidth]{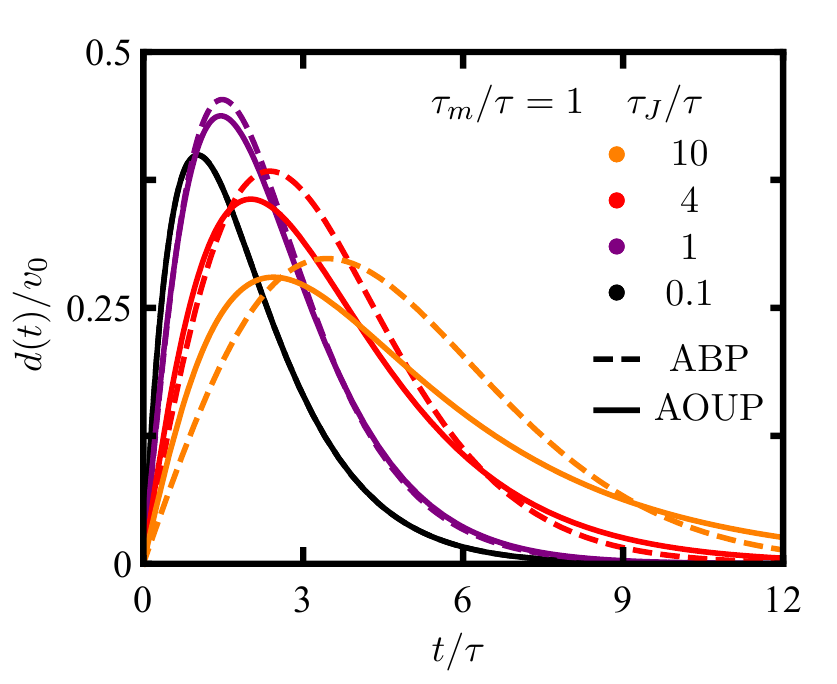}
		\caption{Delay function, $d(t)$, defined in Eq.~\eqref{eq:delay_function},  shown in the same style and for the same parameters as in Fig.~\eqref{fig:velocity_correlation}.}
		\label{fig:delay_function}
	\end{figure}

 Next we consider the delay function between the velocity and orientation of the inertial active particle, defined as~\cite{scholz2018inertial, sprenger2021time} 
	\begin{equation}
	d(t) =  \mean{\vec{v}(t) \cdot \vec{n}(0)} - \mean{\vec{v}(0) \cdot \vec{n}(t)}\,.
	\label{eq:delay_function}
	\end{equation}	
	The two required correlation functions are given by Eq.~\eqref{eq:vini} for the inertial AOUP.
 The inertial delay $d(t)$ has been introduced in Ref.~\cite{scholz2018inertial} as one of the main dynamical effects characterizing ABP with inertia: the velocity $\vec{v}$ tends to lag behind the self-propulsion $\vec{n}$ at a typical delay time.

We see in Fig.~\ref{fig:delay_function} that our model also provides an accurate qualitative picture of effect of rotational inertia on the delay function, regarding both the maximal delay and the characteristic duration of this effect.
Comparing the prediction to the inertial ABP, we find two crossover regimes for large moment of inertia: the inertial AOUP predicts a stronger delay at both short and long times.

	Another benefit of our closed AOUP result is that the total inertial delay $d_\text{tot}:=\int\mathrm{d}t\,d(t)$, i.e., the time integral of Eq.~\eqref{eq:delay_function}, can be determined in the compact form
	\begin{equation}
		d_\text{tot} =  \frac{2 v_0 \tau_m (\tau \tau_\chi +  \tau \tau_m  +\tau_\chi \tau_m ) }{  (\tau + \tau_m) (\tau_\chi + \tau_m)} \,,
	\end{equation}
	which immediately reveals that the delay effect is enhanced by increasing either of the relevant time scales of inertial active motion, given by Eq.~\eqref{eq:ABPparsTAU}.

	\subsection{Positional correlation functions}

    The conditional mean displacement for a given initial value $\vec{n}_0=\vec{n}(0)$ of the self-propulsion vector can be calculated for an inertial AOUP as	
	\begin{align} \label{eq:mean_displacement_aoup}
	\mean{\Delta \vec{r}(t) \vert \vec{n}_0}  = & \mean{ \vec{v} \vert \vec{n}_0 } \tau_m \Big( 1 - e^{-t / \tau_m } \Big) + v_0 \vec{n}_0 \big( \tau + \tau_\chi \big) \nonumber \\
    & + v_0 \vec{n}_0 \bigg( \frac{\tau_m e^{-t/\tau_m} (\tau  \tau_m-\tau  \tau_\chi +\tau_m \tau_\chi )}{(\tau -\tau_m) (\tau_m-\tau_\chi )}    \nonumber\\
& -\frac{\tau ^3 e^{-t/\tau}}{(\tau -\tau_m) (\tau -\tau_\chi )} -\frac{\tau_\chi ^3 e^{-t/\tau_\chi}}{(\tau_\chi -\tau ) (\tau_\chi -\tau_m)} \bigg),  
\end{align}
where we have used the initial condition $\vec{\chi}_0 = \mean{ \vec{\chi} \vert \vec{n}_0 }=\vec{n}_0/\sqrt{\tau}$ for the auxiliary process $\vec{\chi}$ (see Eq.~\eqref{eq:mean_chi}) and
the initial velocity $\mean{ \vec{v} \vert \vec{n}_0 }$ follows from Eq.~\eqref{eq:condvn}.
For $t\to\infty$ we find the persistence length
	\begin{equation}
	L_\text{p} =  \mean{ \vec{v} \vert \vec{n}_0 } \tau_m + v_0 \vec{n}_0 \big( \tau + \tau_\chi \big)\,,
	\label{eq:MDAOUP}
	\end{equation}
which has the same form as that of an inertial ABP.

	\begin{figure}
		\includegraphics[width=0.9\columnwidth]{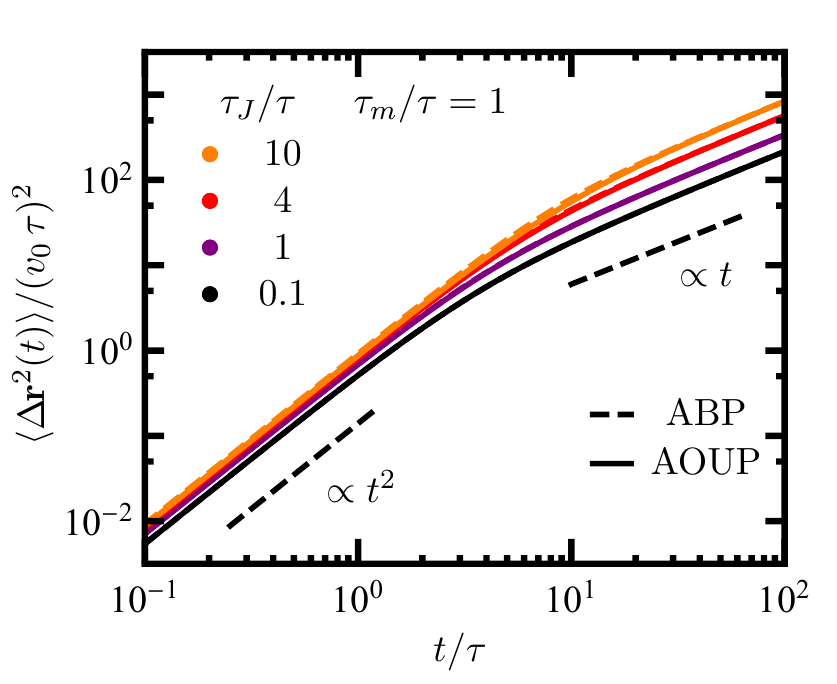}
		\caption{Mean-square displacement, $\mean{ \Delta \vec{r}^2(t) }$, shown in the same style (mind the logarithmic scales) and for the same parameters as in Fig.~\eqref{fig:velocity_correlation}.
		The curves for the two models cannot be distinguished here.
		}
		\label{fig:msd}
	\end{figure}

Moreover, the mean-square displacement $\text{MSD}(t)$ can be expressed as 	
\begin{align}
	\mean{ \Delta \vec{r}^2(t) } = & 4 D_L t + 2 \big(  \mean{\vec{v}(t) \cdot \vec{v}(0)} - \mean{\vec{v}^2} \big) \tau_m^2  \\
	& -  \frac{2 v_0^2}{ \tau - \tau_\chi} \bigg( \tau^3 (1 - e^{-t/\tau}) - \tau_\chi^3 (1-e^{-t/\tau_\chi})  \bigg) \,, \nonumber
	\end{align}
where the velocity correlation $\mean{\vec{v}(t) \cdot \vec{v}(0)}$ and the mean-square velocity $\mean{\vec{v}^2}$ are given by Eq.~\eqref{eq:velocitycorrelation} and Eq.~\eqref{eq:msv}, respectively.
The long-time diffusion coefficient 
\begin{equation} \label{eq:Long_time_diffusion_coefficient_aoup}
	D_L = D_t +\frac{v_{0}^{2}}{2 } \big(\tau + \tau_\chi \big) 
	\end{equation}
	is in full agreement with that of an inertial ABP.
	As shown in Fig.~\ref{fig:msd} the mean-square displacement $\text{MSD}(t)$ of inertial AOUPs agrees fairly well with the ABP result also at intermediate times.

    \section{Conclusions} \label{sec_CON}

	In this paper, we have generalized the inertial active Ornstein-Uhlenbeck particle (AOUP) model to account for translational and, in particular, for rotational inertia in two spatial dimensions.
	The inertial AOUP model introduced in this paper goes beyond mapping the rotational inertia onto an effective rotational diffusion coefficient~\cite{lisin2022motion}
 by incorporating a second characteristic time scale
 (in addition to the one related to the inverse rotational diffusion coefficient),
 which we have demonstrated to be the crucial ingredient for describing the proper long-time behavior.
    As such, our model matches both the small- and long-time regime with the inertial ABP model and thus represents a suitable alternative, which allows to determine closed analytical predictions for dynamical correlations. 
    Indeed, the agreement between inertial ABP and AOUP models has been certified by comparing velocity correlations, the delay function and the mean-square displacement. 
	 For small or moderate moment of inertia, we have found similar predictions of these two models at all times, while small deviations only occur at intermediate times for large moment of inertia. 
	 In general, the effect of increasing rotational inertia is qualitatively captured well by the inertial AOUP model. 
	 In conclusion we have introduced and validated a Gaussian model to describe inertial active matter, which can be considered as an alternative to the inertial ABP model.

	In analogy with the overdamped AOUP model, we expect that the inertial AOUP model presented here will offer an intriguing platform to provide analytic insight into various phenomena exhibited by active particles governed by both translational and rotational inertia.
	Most notably, future studies could focus on the generalization of effective-equilibrium theories with the inertial AOUP as a starting point. 
	The extension of the unified colored noise approximation (UCNA)~\cite{jung1987dynamical, hanggi1995colored, maggi2015multidimensional, caprini2021correlated} or Fox approach~\cite{fox1986, fox1986b, sharma2017escape, wittmann2017effective} will helpful to understand the behavior of inertial active particles in the presence of interactions.
 
	While, recently, it was shown that rotational inertia is able to promote phase separation~\cite{caprini2022role} in purely repulsive systems, further interesting questions remain to be addressed at the collective level.
  For example, the effect of rotational inertia on the (continuous or discontinuous) nature of MIPS~\cite{su2021inertia} or on the kinetic temperature difference between high- and low-density phases~\cite{mandal2019motility} is still unexplored.
 More generally, it would interesting to shed light on the effect of inertia on the recent micro-phase separation observed in field theories~\cite{tjhung2018cluster, fausti2021capillary} and overdamped particle-based simulations of repulsive ABPs~\cite{caporusso2020motility,shi2020self} or dumbbells~\cite{tung2016micro}.
To this end, it will be insightful to apply effective interactions~\cite{farage2015effective,wittmann2016active} or hydrodynamics~\cite{marconi2021hydrodynamics, omar2022mechanical} and mean-field methods~\cite{speck2015dynamical}, to obtain theoretical predictions that take advantage of the intrinsic simplicity of the inertial AOUP model.

	\appendix

	\section{Results for an inertial ABP} \label{app_ABP}
	
	For reference, we summarize here the essential analytic results of the inertial ABP model. 
	Using methods of stochastic integration, we obtain the orientational correlation function in the steady state as
	\begin{align} \label{eq:C(t)_abp}
		\mean{\vec{n}(t)\cdot\vec{n}(0)} = e^{-D_{r}\big(t - \tau_J (1-e^{-t / \tau_J }) \big)}.
	\end{align}
	A characteristic orientational persistence time $\tau_\text{p}$ can  be determined as
	\begin{equation} \label{eq:persistence_time_abp}
		\tau_\text{p} = \int_{0}^{\infty} \!  \mean{\vec{n}(t)\cdot\vec{n}(0)} \mathrm{d}t = \tau_J e^{\mathcal{J}}\mathcal{J}^{-\mathcal{J}} \, \Gamma(\mathcal{J},0,\mathcal{J})
	\end{equation}
	with the reduced moment of inertia $\mathcal{J}:= \tau_J/\tau$.
	
	Similarly, the translational velocity correlation function can be computed as
	\begin{equation} \label{eq:Z(t)_abp}
		\mean{\vec{v}(t) \cdot \vec{v}(0)} = \frac{2 D_t}{ \tau_m } \, e^{-t / \tau_m} + \frac{v_{0}}{2} \big( \mean{\vec{v}(t)\cdot\vec{n}(0)}  + \mean{\vec{v}(0)\cdot\vec{n}(t)} \big),
	\end{equation}
	as well as the delay function 
	\begin{equation} \label{eq:d(t)_abp}
		d(t) = \mean{\vec{v}(t)\cdot\vec{n}(0) }-\langle\vec{v}(0)\cdot \vec{n}(t) \rangle
	\end{equation}
	with
	\begin{alignat}{1} 
		\mean{\vec{v}(t)\cdot\vec{n}(0)} = & v_0 \frac{\tau_J}{\tau_m}  e^{\mathcal{J}} \Big(  \mathcal{J}^{-\Omega_{-}} \Gamma( \Omega_{-}, \mathcal{J} e^{-t/\tau_J}, \mathcal{J}) \nonumber \\
		& + \mathcal{J}^{-\Omega_{+}} \Gamma(\Omega_{+}, 0, \mathcal{J}) \Big) e^{-t / \tau_m} , \label{eq:V(t)n(0)_abp} \\
		\mean{\vec{v}(0)\cdot\vec{n}(t)} = &  v_0 \frac{\tau_J}{\tau_m} e^{\mathcal{J}} \mathcal{J}^{-\Omega_{+}} \Gamma( \Omega_{+}, 0, \mathcal{J} e^{-t / \tau_J}) e^{t / \tau_m} \label{eq:V(0)n(t)_abp}
	\end{alignat}
	and $\Omega_\pm  = \tau_J/\tau \pm \tau_J /\tau_m  $.

	Next, we address the mean displacement $\mean{\Delta \vec{r}(t)\vert \vec{n}_0}$ at prescribed initial orientation $\vec{n}_0$, which reads
	\begin{align} \label{eq:mean_displacement_abp}
		\mean{\Delta \vec{r}(t)\vert \vec{n}_0}  = & \mean{\vec{v}\vert\vec{n}_0} \tau_m \Big( 1 - e^{-t / \tau_m } \Big) \\
		&+ \frac{v_{0}}{D_{r}} \mathcal{J} e^{\mathcal{J}} \Big(  \mathcal{J}^{-\mathcal{J}} \Gamma( \mathcal{J}, \mathcal{J} e^{-t / \tau_J}, \mathcal{J}) \nonumber \\
		&+ \mathcal{J}^{-\Omega_{-}} \Gamma( \Omega_{-}, \mathcal{J} e^{-t / \tau_J}, \mathcal{J}) e^{- t / \tau_m} \Big) \uvec{n}_0 \nonumber
	\end{align}
	with the mean initial velocity
	\begin{equation}
		\mean{ \vec{v}\vert\vec{n}_0 } = v_0 \frac{\tau_J}{\tau_m} e^{\Omega} \Omega^{-\Omega_{+}} \Gamma(\Omega_{+}, 0, \mathcal{J}) \uvec{n}_0
	\end{equation} 
	at given $\vec{n}_0$.
	Thus, the long-time limit of Eq.~\eqref{eq:mean_displacement_abp} yields the persistence length
	\begin{equation}
	L_\text{p} =  \mean{ \vec{v} \vert \vec{n}_0 } \tau_m + v_0 \vec{n}_0 \tau_\text{p} \,,
	\end{equation}
	which has the same form as Eq.~\eqref{eq:MDAOUP}, while the required expression for $\mean{ \vec{v} \vert \vec{n}_0 }$ differs.

	Last, the mean-square-displacement (MSD) is given by 
	\begin{align}
	\mean{ \Delta \vec{r}^2(t) }  =  &   4 D_L t  + 2 \big(  \mean{\vec{v}(t) \cdot \vec{v}(0)} - \mean{\vec{v}^2} \big) \tau_m^2   \label{eq:MSD_abp} \\
	& + 2 v_{0}^{2} \tau_J^{2} \frac{e^{  \mathcal{J} }}{\mathcal{J}^2} \Bigg(  \pFq{2}{2}{\mathcal{J},\mathcal{J}}{\mathcal{J}+1,\mathcal{J}+1}{-  \mathcal{J} } \nonumber \\
	&-  \pFq{2}{2}{\mathcal{J},\mathcal{J}}{\mathcal{J}+1,\mathcal{J}+1}{- \mathcal{J} e^{- t/ \tau_J }} e^{- t/\tau }  \Bigg) \nonumber
	\end{align}
	with the long-time diffusion coefficient
	\begin{equation} \label{eq:Long_time_diffusion_coefficient_abp}
		D_L = D_t +\frac{v_{0}^{2}}{2 } \tau_\text{p}
	\end{equation}
	and the generalized hypergeometric function  ${}_pF_q$.

	\section{Stationary probability distribution for the inertial AOUP model}\label{aoup:steady_states}
	
	In this Appendix, we derive the stationary probability distribution $\fct{P}(\vec{v}, \vec{n}, \vec{\chi})$  for the AOUP model with rotational and translational inertia.
	First, we note that Eqs.~\eqref{eq:langevin_v},~\eqref{eq:langevin_n} and~\eqref{eq:langevin_chi}, can be written in the form 
\begin{equation}
	\dot{\vec{w}} = - \vec{\mathcal{A}} \, \vec{w} + \vec{\sigma} \, \vec{\eta} \,,
\end{equation}
	where $\vec{\mathcal{A}}$ and $\vec{\sigma}$ are the drift and the noise matrices, respectively, $\vec{w}$ the vector of dynamical variables, and $\vec{\eta}$ a white noise vector with unit-variance. 
	The stationary probability distribution of this system is a multivariate Gaussian of the form
\begin{equation}
	\fct{P}(\vec{w}) \propto \exp{\big( - \vec{w}^{T} \,  \vec{\mathcal{C}}^{-1} \, \vec{w}  \big)}\,, 
\end{equation}	
	where $\vec{\mathcal{C}}^{-1}$ is the inverse of the correlation matrix $\vec{\mathcal{C}}$ to be determined by solving the following matrix equation
\begin{equation}
	\label{eq:matrix_equation}
	\vec{\mathcal{A}} \,
	\vec{\mathcal{C}}+\vec{\mathcal{C}} \, \vec{\mathcal{A}}^{T} =  \vec{\sigma} \vec{\sigma}^{T}\,.
\end{equation}
	Here, $\vec{\mathcal{A}}^T$ and $\vec{\sigma}^{T}$ the transpose of drift and noise matrix, respectively.
	
	Applying this general approach for $\vec{w}= (\vec{v}, \vec{n}, \vec{\chi})$, we obtain
	{\allowdisplaybreaks
	\begin{align}
		 \fct{P}(&\vec{v}, \vec{n}, \vec{\chi}) \propto \exp{\left( - \frac{\vec{v}^2}{2} \mathcal{C}^{-1}_{\vec{v}\vec{v}} - \frac{\vec{n}^2}{2} \mathcal{C}^{-1}_{\vec{n}\vec{n}}  - \frac{\vec{\chi}^2}{2} \mathcal{C}^{-1}_{\vec{\chi}\vec{\chi}} \right)} \nonumber \\
		& \times \exp{ \Big(- \vec{v}\cdot\vec{n} \, \mathcal{C}^{-1}_{\vec{v} \vec{n}} - \vec{v}\cdot\vec{\chi} \, \mathcal{C}^{-1}_{\vec{v} \vec{\chi}} - \vec{n}\cdot\vec{\chi} \, \mathcal{C}^{-1}_{\vec{n} \vec{\chi}}  \Big)}, \label{eq:probability_v_n_chi}
	\end{align}}
	where 
	{\allowdisplaybreaks
	\begin{widetext}
		\begin{subequations}
			\begin{align}
				\mathcal{C}_{\vec{v}\vec{v}}^{-1}       = & \bigg( \frac{D_t}{\tau_m} + \frac{v_0^2 \tau_m^3 (\tau + \tau_\chi)}{2 (\tau +\tau_m)^2 (\tau_\chi + \tau_m)^2} \bigg)^{-1}, \\
				\mathcal{C}_{\vec{n}\vec{n}}^{-1}       = &  \frac{\tau + \tau_\chi}{\tau} \bigg( \frac{2D_t}{\tau_m} + v_0^2 \frac{\tau_\chi^2 \tau_m^2 + \tau^2 (\tau_\chi + \tau_m)^2+ \tau \tau_m \tau_\chi(2 \tau_m + 3\tau_\chi) }{ (\tau + \tau_m) (\tau + \tau_\chi )(\tau_\chi + \tau_m)^2 } \bigg) \bigg( \frac{D_t}{\tau_m} + \frac{v_0^2 \tau_m^3 (\tau + \tau_\chi)}{2 (\tau +\tau_m)^2 (\tau_\chi + \tau_m)^2} \bigg)^{-1}, \\
				\mathcal{C}_{\vec{\chi}\vec{\chi}}^{-1} = & 2 \tau_\chi +  \frac{v_0^2 \tau \tau_\chi^2 \tau_m^2}{(\tau +\tau_m)^2 (\tau_\chi + \tau_m)^2} \bigg( \frac{D_t}{\tau_m} + \frac{v_0^2 \tau_m^3 (\tau + \tau_\chi)}{2 (\tau +\tau_m)^2 (\tau_\chi + \tau_m)^2} \bigg)^{-1}, \\
				\mathcal{C}_{\vec{v}\vec{n}}^{-1}       = & -  \frac{v_0}{\tau + \tau_m} \bigg( \tau + \frac{2 \tau_\chi \tau_m}{\tau_\chi + \tau_m} \bigg) \bigg( \frac{D_t}{\tau_m} + \frac{v_0^2 \tau_m^3 (\tau + \tau_\chi)}{2 (\tau +\tau_m)^2 (\tau_\chi + \tau_m)^2} \bigg)^{-1},\\
				\mathcal{C}_{\vec{v}\vec{\chi}}^{-1}    = &   \frac{v_0 \sqrt{\tau} \tau_\chi \tau_m}{(\tau + \tau_m) (\tau_\chi + \tau_m)} \bigg( \frac{D_t}{\tau_m} + \frac{v_0^2 \tau_m^3 (\tau + \tau_\chi)}{2 (\tau +\tau_m)^2 (\tau_\chi + \tau_m)^2} \bigg)^{-1}, \\
				\mathcal{C}_{\vec{n}\vec{\chi}}^{-1}    = & - \frac{ \tau_\chi}{\sqrt{\tau}} \bigg( \frac{2D_t}{\tau_m} + \frac{v_0^2 \tau_m}{(\tau + \tau_m) (\tau_\chi +\tau_m)} \bigg( \tau + \frac{ \tau_\chi \tau_m}{\tau_\chi + \tau_m} \bigg) \bigg) \bigg( \frac{D_t}{\tau_m} + \frac{v_0^2 \tau_m^3 (\tau + \tau_\chi)}{2 (\tau +\tau_m)^2 (\tau_\chi + \tau_m)^2} \bigg)^{-1}\,. 
			\end{align}
		\end{subequations}
	\end{widetext}}
	The stationary probability distribution $\fct{P}(\vec{v}, \vec{n}, \vec{\chi})$ (Eq.~\eqref{eq:probability_v_n_chi}) can be rewritten as
	\begin{equation}
		\fct{P}(\vec{v}, \vec{n}, \vec{\chi}) = \fct{P}(\vec{v}\vert\vec{n},\vec{\chi}) \fct{P}(\vec{n}, \vec{\chi})\,,
	\end{equation}
	where $\fct{P}(\vec{n}, \vec{\chi})$ is the reduced probability describing the active self-propulsion and the $\fct{P}(\vec{v}\vert\vec{n},\vec{\chi})$ defines the conditional probability to find a particle at a velocity $\vec{v}$ with prescribed $\vec{n}$ and $\vec{\chi}$
	{\allowdisplaybreaks
	\begin{subequations}
		\begin{align}
			\fct{P}(\vec{v}\vert\vec{n}, \vec{\chi}) \propto & \exp \bigg( - \frac{\big( \vec{v} - \mean{\vec{v}\vert\vec{n},\vec{\chi}} \big)^2}{ \, \sigma(\vec{v}\vert\vec{n},\vec{\chi}) }   \bigg) , \\  
			\mean{\vec{v}\vert\vec{n},\vec{\chi}} = &  \frac{v_0 ( \tau \tau_m  +  2\tau_m \tau_\chi + \tau \tau_\chi)}{(\tau + \tau_m)(\tau_\chi+\tau_m)} \, \vec{n} \nonumber \\
			& - \frac{v_0 \sqrt{\tau} \tau_m \tau_\chi }{(\tau + \tau_m)(\tau_\chi+\tau_m)} \, \vec{\chi}, \\
			\sigma(\vec{v}\vert\vec{n},\vec{\chi}) = &  \frac{2 D_t}{\tau_m} + \frac{v_0^2 \tau_m^3 (\tau + \tau_\chi)}{ (\tau +\tau_m)^2 (\tau_\chi + \tau_m)^2}\,. 
		\end{align}
	\end{subequations}}
	The latter distribution fluctuates around the conditional average $\mean{\vec{v} \vert \vec{n},\vec{\chi}}$ of $\vec{v}$ at given $\vec{n}$ and $\vec{\chi}$ with its corresponding variance $\sigma(\vec{v} \vert \vec{n},\vec{\chi})$.
Integration over the auxiliary process $\vec{\chi}$ yields the results stated and discussed in Sec.~\ref{sec:Pvn}
 
	In a similar way, the reduced probability $\fct{P}(\vec{n}, \vec{\chi})$ can be expressed as
	\begin{equation}
		\fct{P}(\vec{n}, \vec{\chi}) = \fct{P}(\vec{n}\vert \vec{\chi}) \fct{P}(\vec{\chi}) \,
	\end{equation}
	with
	\begin{subequations}
		\begin{align}
			\fct{P}(\vec{n}\vert \vec{\chi}) & \propto \exp{\bigg( - \frac{\big( \vec{n} - \mean{\vec{n}\vert\vec{\chi}} \big)^2}{\sigma(\vec{n}\vert\vec{\chi})  } \bigg)}, \\  
			\mean{\vec{n}\vert\vec{\chi}} & = \frac{\sqrt{\tau} \tau_\chi }{\tau+ \tau_\chi} \vec{\chi}, \\
			\sigma(\vec{n}\vert\vec{\chi}) & = \frac{\tau}{\tau + \tau_\chi} \,
		\end{align}
	\end{subequations}
	and
	\begin{subequations}
		\begin{align}
			\fct{P}(\vec{\chi}) &\propto \exp{\left( - \frac{\vec{\chi}^2}{ \mean{\vec{\chi}^2}} \right)}, \\  
			\mean{\vec{\chi}^2} & = \frac{\tau+ \tau_\chi}{\tau \tau_\chi} \,,
		\end{align}
	\end{subequations}
	or alternatively
	\begin{equation}
		\fct{P}(\vec{n}, \vec{\chi}) = \fct{P}(\vec{\chi}\vert \vec{n}) \fct{P}(\vec{n}) \,
	\end{equation}
	with
	\begin{subequations}
		\begin{align}
			\fct{P}(\vec{\chi}\vert \vec{n}) & \propto \exp{\bigg( - \frac{\big( \vec{\chi} - \mean{\vec{\chi}\vert\vec{n}} \big)^2}{ \sigma(\vec{\chi} \vert \vec{n}) } \bigg)}, \\  
			\mean{\vec{\chi}\vert\vec{n}} & =  \vec{n}/\sqrt{\tau}, \label{eq:mean_chi} \\
			\sigma(\vec{\chi} \vert \vec{n}) & = 1/\tau_\chi \,
		\end{align}
	\end{subequations}
	and
	\begin{equation}
		\fct{P}(\vec{n}) \propto \exp{ \big( - \vec{n}^2 \big)} \,.
	\end{equation}	
	The distribution $\fct{P}(\vec{v}, \vec{n})$ (see Eq.\eqref{eq:probability_v_n}) can be derived via integration of the full probability density $\fct{P}(\vec{v}, \vec{n}, \vec{\chi})$  (see Eq.\eqref{eq:probability_v_n_chi}) with respect to $\vec{\chi}$.

	\acknowledgments

LC acknowledges support from the Alexander Von Humboldt foundation, while
HL and RW acknowledge support by the Deutsche Forschungsgemein\-schaft (DFG) through the SPP 2265, under grant numbers LO 418/25-1 (HL) and WI 5527/1-1 (RW).

\bibliographystyle{apsrev4-1}
\bibliography{bib}
	
\end{document}